\documentclass[11pt,nofootinbib,tightenlines,superscriptaddress]{revtex4}
\usepackage{amsmath,amssymb}
\usepackage{bbold}
\usepackage{cleveref}
\usepackage[dvipsnames]{xcolor}
\usepackage{graphicx}
\usepackage{boldline,multirow}
\begin{document}
\title{Generalised scalar-tensor theories and self-tuning}
\author{Edmund J. Copeland}
\affiliation{School of Physics and Astronomy, 
University of Nottingham, Nottingham NG7 2RD, UK} 
\affiliation{Nottingham Centre of Gravity, University of Nottingham, Nottingham NG7 2RD, UK}
\author{Sukhraj Ghataore}
\affiliation{School of Physics and Astronomy, 
University of Nottingham, Nottingham NG7 2RD, UK} 
\affiliation{Nottingham Centre of Gravity, University of Nottingham, Nottingham NG7 2RD, UK}
\author{Florian Niedermann}
\affiliation{Nordita, KTH Royal Institute of Technology and Stockholm University\\
Hannes Alfv\'ens v\"ag 12, SE-106 91 Stockholm, Sweden}
\author{Antonio Padilla}
\affiliation{School of Physics and Astronomy, 
University of Nottingham, Nottingham NG7 2RD, UK} 
\affiliation{Nottingham Centre of Gravity, University of Nottingham, Nottingham NG7 2RD, UK}

\newcommand{\tony}[1]{{\color{red} [[tony: #1]]}}
\date{\today}

\begin{abstract}
    We explore a family of generalised scalar-tensor theories that exhibit self-tuning to low scale anti de Sitter vacua, even in the presence of a large cosmological constant.  We are able to examine  the linearised fluctuations about these vacua and compute the corresponding amplitude.  Thanks to a subtle interplay between a weak scalar coupling and a low scalar mass,  it is possible to exhibit self-tuning and compatibility with solar system tests of gravity without resorting to non-linearities and unreliable screening mechanisms.  The weakness of the scalar coupling and the correspondingly slow response to vacuum energy phase transitions may present some interesting possibilities for connecting early universe inflation to the cancellation of vacuum energy. 
\end{abstract}
\maketitle
\newcommand{\Ac}{\mathcal{A}}
\newcommand{\Wc}{\mathcal{W}}
\newcommand{\Lc}{\mathcal{L}}
\newcommand{\Kc}{\mathcal{K}}
\newcommand{\Tr}{\mathrm{Tr}}
\newcommand{\Uc}{\mathcal{U}}
\newcommand{\Ec}{\mathcal{E}}
\newcommand{\GB}{{\mathcal{G}}}
\newcommand{\BA}{\Bar{\Ac}}
\newcommand{\BK}{\Bar{\Kc}}
\newcommand{\BRingo}{\Bar{V}_r}
\newcommand{\meff}{m_g^2}
\newcommand{\Bnab}{\Bar{\nabla}}
\newcommand{\Bsq}{\Bar{\square}}
\newcommand{\RC}{\mathcal{R}}
\section{Introduction}
The cosmological constant problem has haunted theoretical physics ever since Pauli calculated the effect of the zero-point energy of the electron on the curvature of spacetime, declaring that the universe would "not even reach to the moon" \cite{Enz}. The situation hasn't really improved in the century that followed. When we apply standard quantum field theory methods, radiative corrections to the vacuum energy are extremely sensitive to ultra-violet physics,  scaling like the fourth power of the cut-off.  Vacuum energy gravitates just like a cosmological constant presenting a major issue for cosmology. The observed value of the cosmological constant lies sixty orders of magnitude below the expected value predicted from vacuum energy calculations with a TeV cut-off,  set by the scale of modern collider experiments. If we push the cut-off all the way up to the Planck scale, the discrepancy extends to 120 orders of magnitude.  For reviews of the cosmological constant problem, see \cite{weinberg1989cosmological,Polchinski:2006gy,martin2012everything,Burgess:2013ara, padilla2015lectures}.

One approach to the cosmological constant problem is to invoke some sort of \textit{self-tuning} or \textit{self-adjustment} mechanism. This assumes the existence of additional fields that adjust their value in order to shield the spacetime from a large underlying vacuum energy.  In other words, the self adjusting fields respond to changes in the vacuum energy, leaving the spacetime geometry relatively unaffected.  In his  seminal review article \cite{weinberg1989cosmological}, Weinberg famously developed a no-go theorem ruling out a phenomenologically viable model of self-tuning.  As with any no-go theorem,  there were some key underlying assumptions, including the assumption of a local kinetic structure in the four dimensional effective theory, and translational invariance of the vacuum.  The latter condition was relaxed for so-called Fab Four theories \cite{charmousis2011self,charmousis2012general,Copeland:2012qf}, evading the clutches of Weinberg's theorem and giving rise to self-tuning Minkowski vacua even in the presence of a very large vacuum energy. However, all Fab Four  theories necessarily contained a light scalar coupled to matter with gravitational strength, potentially giving rise to  fifth forces and eventually being ruled out by multi-messenger probes of neutron star mergers \cite{Creminelli:2017sry,Sakstein:2017xjx,Ezquiaga:2017ekz,Bordin:2020fww}. Self-tuning can also be realised in higher dimensional settings (see e.g. \cite{Arkani-Hamed:2000hpr, Charmousis:2017rof,Aghababaie:2003wz,Burgess:2011va,Dvali:2000xg,Dvali:2007kt}) although such scenarios are often beset with problems such as  hidden fine tunings, ghost-like instabilities and bad phenomenology \cite{Dubovsky:2002jm,Hassan:2010ys,Niedermann:2014bqa,Eglseer:2015xla,Niedermann:2015via,Niedermann:2015vbk}. 

These problems prompted two of us \cite{niedermann2017gravitational} to expand on Weinberg's no-go theorem,  dropping the assumption of translation invariance but bringing in other considerations such as stability of the vacuum and observational constraints. The idea was to use K\"{a}ll\'{e}n-Lehmann spectral representations of exchange amplitudes for conserved gravitational sources to explore a very general class of  self adjusting theories.  By assuming unitarity, metastability of the self-tuning vacuum,  and compatibility with solar system tests of gravity at leading order in perturbations \cite{will2014confrontation}, self-tuning on Minkowski and de Sitter vacua was shown to be impossible. However, the analysis did leave room for two interesting loopholes. The first of these is realised by vacuum energy sequestering \cite{Kaloper:2013zca, Kaloper:2014dqa,Kaloper:2014fca,Kaloper:2015jra,Kaloper:2016yfa,Kaloper:2016jsd,DAmico:2017ngr,Padilla:2018hvp,Kaloper:2018kma,Coltman:2019mql,El-Menoufi:2019qva,Lawrence:2020djs,Blanco:2020ipk}, where the free field propagators do not take the canonical form (for sequestering they take the form of decapitated propagators \cite{Adams:2002ft}). The second loophole involves self-tuning on anti de Sitter vacua. In other words, the no-go results of  \cite{niedermann2017gravitational} did not exclude the possibility that there are self adjusting field theories for which the vacuum is low scale anti de Sitter, even in the presence of a large vacuum energy (of any sign).  It is this anti de Sitter loophole that we would like to explore further in this paper. 

We identify a generalised class of Fab Four theories, where the self-tuning vacuum is stable anti de Sitter, and linearised perturbations about the vacuum satisfy solar system constraints on the gravitational force. The anti de Sitter vacuum means that translational invariance is broken by the  metric field as opposed to the vacuum expectation value of the scalar. This should be contrasted with the original Fab Four theory, where the vacuum was Minkowski, and so translational invariance had to be broken by the scalar in order to evade Weinberg's theorem.  The constant vev for the scalar in our new set-up makes it more amenable to an analytic analysis of linear perturbations. Perturbative stability and compatibility with solar system tests require the scalar to be very weakly coupled.  However, it is also very light and can exert a non-trivial effect on long wavelength sources, making self-tuning possible.  

Our generalised Fab Four theories include the four familiar terms - John, Paul, George, and Ringo \cite{charmousis2011self,charmousis2012general,Copeland:2012qf} - although now they are adapted to an anti de Sitter vacuum with curvature $-q^2$, similar to those seen in \cite{fab5}. The curvature scale $q$ is chosen independently of the vacuum energy, and it can be set to lie below the current Hubble scale. Our generalised theory also contains an extension of the dRGT mass term for the graviton \cite{fierz1939relativistic, de2011resummation,hassan2012resolving},  yielding a scalar-graviton potential that is ghost free and similarly adapted to vanish on the bespoke anti de Sitter vacuum. 

As a canonical example, we note that there exists  a simple self-tuning theory where the graviton mass actually vanishes, described by the following action 
\begin{equation} \label{canact}
S=\int d^4 x \sqrt{-g} \left[\frac{M_{pl}^2}{2} R
-\frac12 g^{\mu\nu} \nabla_\mu\phi \nabla_\nu \phi \right. 
\left.
+\frac{\phi}{\mu} (\mathcal{G}-24 q^4)  \right]+S_m[g_{\mu\nu},  \Psi] ,
\end{equation}
where $S_m$ is the matter action and the Gauss-Bonnet combination, $\GB = R_{\mu \nu \alpha \beta}R^{\mu \nu \alpha \beta} - 4 R_{\mu \nu}R^{\mu \nu} + R^2$.  We have also introduced an explicit mass scale, $\mu$.  Even in the presence of a large vacuum energy, $V_{vac}$, the vacuum is anti de Sitter with curvature $-q^2$. The vacuum energy fixes the constant vev of the scalar field as $\bar \phi=-\frac{\mu}{24 q^4} (  V_{vac}+3 M_{pl}^2 q^2)$. Fluctuations about the vacuum are well behaved, and satisfy
solar system tests provided $|\mu| \gtrsim 1000 q^2/M_{pl}$. 

The focus on anti de Sitter vacua, motivated by the no go results of \cite{niedermann2017gravitational}, might raise phenomenological concerns about describing the current phase of acceleration. However,  the transition to self-tuning vacua is not instantaneous - on the contrary, the weakness of the scalar coupling means it will generically be very slow. Therefore, if dark energy begins to dominate at late times, it could easily give rise to a quasi de Sitter expansion for at least an efold, as required by observation.   As it happens, the canonical example presented above also admits self-tuning to de Sitter vacua. However, in that case, the scalar is tachyonic (consistent with \cite{niedermann2017gravitational}), albeit at a scale much lower than the de Sitter curvature. 

The weakness of the scalar coupling also means that the response to a phase transition in the early universe is expected to be slow. This could be a serious problem for all our models, as it suggests that there could be a long period of accelerated expansion after the QCD phase transition, cooling the universe right down.  However, this early phase of acceleration could potentially be identified with inflation, although to become phenomenologically viable, there would also need to be a period of preheating, raising the temperature of the universe to  beyond the scale of nucleosynthesis. The details of this speculative idea are beyond the scope of the current paper, although they do offer some tantalising prospects for future research. 

The rest of this paper is organised as follows:  in the next section, we write down a generalised family of Fab Four theories, exhibiting self-tuning to AdS or dS vacua.  We also allow the graviton to have a mass. In section \ref{sec:fluc}, we consider fluctuations about the self-tuning vacua  and express the results in terms of the amplitude. We  establish constraints that ensure the system is both self-tuning and able to pass solar system tests of gravity in the relevant limit. The details of the amplitude calculation appear in an appendix. In section \ref{sec:can}, we illustrate the results of the previous section by examining the canonical example of \eqref{canact}. In section \ref{sec:dis}, we make some concluding remarks and speculate on possible future directions. 

\section{Generalised Fab Four theories}
Fab Four theories were introduced in \cite{charmousis2011self,charmousis2012general,Copeland:2012qf} as the most general subset of Horndeski theories \cite{Horn, DGSZ} that exhibited dynamical  self-tuning to Minkowski space, whatever the value of the vacuum energy in any given epoch.  As the name suggests, the theory is built from four distinct terms in the Lagrangian, each vanishing identically for vanishing Ricci curvature.  

Here we consider a generalisation of the Fab Four described by a generic action of the form
\begin{subequations}
 \label{action}
\begin{equation}
S=\int d^4 x \sqrt{-g} \left[\mathcal{L}_j+\mathcal{L}_p+\mathcal{L}_g+\mathcal{L}_r \right. 
\left.-\mathcal{K}(\phi) g^{\mu\nu} \nabla_\mu\phi \nabla_\nu \phi -U_{mg} \right]+S_m[g_{\mu\nu},  \Psi] ,
\end{equation}
where we have the usual Fab Four terms - John (j), Paul (p), George (g) and Ringo (r) - adapted to an anti de Sitter vacuum with  curvature $-q^2$,
\begin{align}
&\mathcal{L}_j = V_j(\phi) (G^{\mu\nu}-3q^2 g^{\mu\nu}) \nabla_\mu\phi \nabla_\nu \phi , \\ 
&\mathcal{L}_p = V_p(\phi) (P^{\mu\nu\alpha\beta}  +2q^2  g^{\mu [\alpha} g^{\beta]\nu}) \nabla_\mu\phi \nabla_\alpha \phi \nabla_\nu \nabla_\beta \phi , \quad~   \\
&\mathcal{L}_g = V_g(\phi) (R+12 q^2) , \\
&\mathcal{L}_r =V_r(\phi) (\mathcal{G}-24 q^4) .
\end{align}
\end{subequations}
These four terms are obtained from the original Fab Four \cite{charmousis2011self,charmousis2012general,Copeland:2012qf} by replacing the curvature terms in the Lagrangian with a shifted curvature, $R_{\mu\nu\alpha\beta} \to R_{\mu\nu\alpha\beta} +q^2 (g_{\mu\alpha}g_{\nu\beta}-g_{\mu\beta}g_{\nu\alpha})$.   In addition to the Einstein tensor, the Ricci scalar and the Gauss-Bonnet combination, $\GB$, the Fab Four depends on the double dual of the Riemann tensor, $P^{\mu \nu}{}_{\alpha \beta}  = -\frac{1}{4}\delta^{\mu \nu \rho \delta}_{\sigma \lambda \alpha \beta} R^{\sigma \lambda}{}_{\rho \delta} = -R^{\mu \nu}{}_{\alpha \beta} + 2R^\mu_{[\alpha}\delta^\nu_{\beta ]} - 2 R^\nu_{[ \alpha}\delta^\mu_{\beta ]} - R\delta^\mu_{[ \alpha}\delta^\nu_{\beta ]}$. 

 The action \eqref{action} also contains an additional kinetic term for the scalar, proportional to a general function $\mathcal{K}(\phi)$, whose vacuum expectation value can be chosen so that  linearised perturbations satisfy solar system constraints \cite{will2014confrontation}. This term is absent in standard Fab Four models, but it is allowed here since it does not affect the form of the vacuum solution for a constant scalar field.  The potential $U_{mg}$ is a generalisation of the dRGT mass term for the graviton \cite{fierz1939relativistic, de2011resummation,hassan2012resolving},  being ghost free and again adapted to vanish on an anti de Sitter vacuum with  curvature $-q^2$.    Specifically, we have 
\begin{equation}
U_{mg}=\frac14 m^2V_g^2\Big[\Uc(\Sigma)\Big]^{F_E}_{\bar F_E},
\end{equation}
where we introduce the shorthand $\Big[\Uc(\Sigma)\Big]^{F_E}_{\bar F_E}=\mathcal{U}(F_E)-\mathcal{U}(\bar F_E)$ for later convenience.  The corresponding arguments are given implicitly as $(F_E)^2 =\tilde g^{-1}\Bar g= \frac{1}{V_g} g^{-1} \Bar{g}$, $(\bar F_E)^2 = \frac{1}{ V_g} \mathbb{1}$  and are built directly from the Einstein frame metric $\tilde g_{\mu\nu}=V_g g_{\mu\nu}$ and the vacuum anti de Sitter metric, $\Bar{g}_{\mu\nu}$. The underlying potential takes the familiar dRGT form \cite{de2011resummation}
\begin{equation}
    \Uc (\Sigma ) = -4 \big( 12 - 6\,  \Tr \Sigma +  \big( \Tr \Sigma \big)^2  - \Tr \Sigma^2 \big)  .
\end{equation}
Our goal here is to explore the phenomenology of these generalised models. In particular, we would like to ask if they exhibit consistent vacua that self-tune to anti de Sitter (with curvature $-q^2$), and whether linearised perturbations  are free of instabilities and compatible with solar system tests of gravity.  Unlike in the original Fab Four set-up, the breaking of translational invariance on the anti de Sitter vacuum allows self-tuning to occur even for constant scalars, making the linearised analysis far simpler.  The John and Paul terms will not play any role whatsoever in this analysis, either at the level of the background, or at linear order.  For that reason, we drop them from now on.  It may be worth restoring them in a detailed analysis of the cosmological evolution which is beyond the scope of this paper. 

With these truncations, our working action is now given by 
\begin{multline} \label{action1}
S=\int d^4 x \sqrt{-g} \Bigg[\Ac(\phi)^2 (R+12 q^2)+V_r(\phi) (\mathcal{G}-24 q^4) \
\\-\mathcal{K}(\phi) g^{\mu\nu} \nabla_\mu\phi \nabla_\nu \phi  -\frac{m^2}{4} \Ac (\phi )^4\Big[ \Uc (\Sigma ) \Big]^{F_E}_{\frac{1}{A}\mathbb{1}}\Bigg] 
+S_m[g_{\mu\nu},  \Psi] ,
\end{multline}
where we have written the George potential as $V_g(\phi)=\Ac(\phi)^2$, for notational convenience.  This yields  field equations $ \Ec^\mu_\nu=0$ from variation with respect to the metric, and $\Ec_\phi=0$ from variation with respect to the scalar, where
\begin{widetext}
    \begin{multline}
    \label{EFE}
        \Ec^\mu_\nu = 2\big( \nabla^\mu \nabla_\nu - \delta^\mu_\nu \square - G^\mu_\nu \big) \Ac ^2 
        + 12q^2 \Ac ^2 \delta^\mu_\nu 
        + 2\Kc \Big\{ \nabla^\mu  \phi \nabla_\nu \phi - \frac{1}{2} \big( \nabla \phi \big)^2 \delta^\mu_\nu \Big\} \\
        + \frac{m^2}{4} \bigg\{ \Wc^\mu_\nu - \delta^\mu_\nu \Big[ \Uc (\Sigma ) \Big]^{F_E}_{\frac{1}{A}\mathbb{1}} \bigg\} \Ac^4   
        + \big( 8P^{\mu \alpha}{}_{\nu \beta} \nabla_\alpha \nabla^\beta - 24q^4\delta^\mu_\nu \big) V_r
        + T^\mu_\nu ,
    \end{multline}
    \begin{multline}
    \label{ScalarEF}
        \Ec_\phi = 2\Ac \Ac ' \big( R + 12q^2\big) + \Kc ' \big( \nabla \phi \big)^2 + 2\Kc \square \phi 
        - \frac{m^2}{4} \Ac^3 \Ac ' \bigg[ 4\Uc (\Sigma ) - \Tr \bigg( \Sigma \frac{\partial \Uc}{\partial \Sigma} \bigg) \bigg]^{F_E}_{\frac{1}{\Ac}\mathbb{1}} + V_r ' \big[ \GB - 24q^4 \big] .
    \end{multline}
    \end{widetext}
 Here  $\Wc^\mu_\nu  = -4 \big( -6(F_E)^\mu_\nu + 2\Tr (F_E)(F_E)^\mu_\nu - 2(F_E)^\mu_\alpha (F_E)^\alpha_\nu \big)$ is related to the energy momentum tensor for the dRGT mass term,  while $T_{\mu \nu} = -\frac{2}{\sqrt{-\Bar{g}}}\frac{\delta S_m}{\delta \Bar{g}^{\mu \nu}}$ is the energy-momentum tensor  for matter minimally coupled to the metric. Prime denotes differentiation with respect to $\phi$. 

We now consider the self-tuning vacua in this theory, assuming the presence of an arbitrarily large vacuum energy, $T_{\mu\nu}=-V_{vac} g_{\mu\nu}$. It is easy to see that the scalar equation of motion \eqref{ScalarEF} is satisfied for a constant scalar $\bar \phi$ whenever the  metric coincides with the background anti de Sitter vacuum $g_{\mu\nu}=\bar g_{\mu\nu}$, with curvature $-q^2$.  The metric equation of motion \eqref{EFE}, sourced by a large vacuum energy, now fixes the value of the scalar field,  as opposed to the metric, giving
\begin{equation}
\label{SelfTuningEq}
    \frac{V_{vac}}{ 6\BA^2} 
    = m^2 \big( \BA - 1 \big) + q^2 \left(1
    - \frac{4q^2}{ \BA^2}\BRingo \right) ,
\end{equation}
where bar denotes the corresponding potential evaluated at $\bar \phi$.  As we will see in a moment, $\BA$ sets the scale of gravitational interactions on this vacuum and is therefore identified with the Planck scale, $M_{pl} \sim 10^{18}$ GeV.  It is instructive to see what happens in the absence of the mass term and the Ringo term. In this instance, equation \eqref{SelfTuningEq} forces a fine tuning condition on the vacuum energy.  Indeed, the background curvature  should not exceed the curvature scale of the universe today, $q^2 \lesssim H_0^2$, and so the vacuum energy is constrained to be unnaturally small $|V_{vac}| \lesssim M_{Pl}^2 H_0^2$.  
However, the presence of the Ringo term immediately changes the story.  Equation  \eqref{SelfTuningEq} now picks up a scalar dependence which we can use to accommodate a much larger vacuum energy.  We might also imagine that the mass term can be used in a similar way. However, as we will see, this is not the case.  A detailed study of the fluctuations suggests that  the mass term needs to be too small, so Ringo is absolutely crucial. 

\section{Fluctuations about the self-tuning vacuum} \label{sec:fluc}
We now consider the fluctuations about this vacuum. As shown in the appendix, this yields the following expression for the exchange amplitude between conserved sources $\tau_{\mu\nu}$ and $\tau'_{\mu\nu}$
\begin{widetext}
\begin{multline} 
\label{Amp}
    \mathrm{Amp} = -\frac{1}{\BA^2} \int \mathrm{d}^4x \sqrt{-\bar g}\bigg\{ \tau'^{\mu \nu} \frac{1}{\Bsq +2q^2 - \meff} \tau_{\mu \nu} 
     - \frac{\tau '}{2} \bigg[ \frac{1/2}{\Bsq +2q^2 -\meff} + \bigg( \frac{q^2 + \meff /6}{q^2 + \meff/2}\bigg) \frac{1/2}{\Bsq -6q^2 -\meff} \bigg] \tau
    \\
    + \frac{m^2_\phi}{12 \Gamma} \tau ' \frac{1}{\Bsq -  m^2_\phi} \tau \bigg\} ,
\end{multline}
\end{widetext}
where the trace of the source is $\tau=\bar g^{\mu\nu} \tau_{\mu\nu}$ and $\tau'=\bar g^{\mu\nu }\tau'_{\mu\nu}$. We also define the effective mass of the graviton $$\meff \equiv m^2 (3\BA -2), $$ and the effective mass of the scalar
\begin{equation} \label{mphi}
m^2_\phi \equiv \frac{12( \alpha \BA^2 \Gamma - 4q^4 \BRingo ')^2}{\BA^2 \Gamma ( \BK + 6\alpha^2 \BA^2 - 24q^2\alpha \BRingo ' ) + 48 q^6 \BRingo '^2},
\end{equation}
where $\alpha \equiv \frac{\BA '}{\BA}$ and $\Gamma \equiv 2q^2 +\meff$.

The first line in \eqref{Amp} is the amplitude for a massive graviton of mass $\meff$ in a maximally symmetric space with curvature $-q^2$.  To avoid a helicity-2 ghost, we require $\BA^2>0$.  If $\meff=0$, there is nothing more to add as the helicity- 1 and helicity-0 modes are absent. However, if $\meff \neq 0$, the helicity- 1 and helicity-0 modes return, and the absence of a corresponding ghost  requires $\meff>0$ and $\Gamma>0$ respectively\footnote{In \cite{niedermann2017gravitational}, only the condition $\Gamma>0$ is required. However, as emphasized in \cite{Claudia}, the condition $\meff>0$ is also required to ensure the stability of the helicity-1 mode.}.  The second line in  \eqref{Amp} is just the amplitude for exchange of a scalar.  This is ghost-free provided $m^2_\phi/\Gamma \geq 0$, and tachyon-free provided $m_\phi^2\geq 0$. In principle, for the case of a massless graviton, we could tolerate a tachyonic instability that was sufficiently slow compared with the current age of the universe. In other words, we might be able to relax the tachyon condition to $m_\phi^2 \gtrsim -H_0^2$, where $H_0$ is the current Hubble scale.

The full amplitude \eqref{Amp} exhibits self-tuning, as long as $m_\phi^2$ and $q^2+\meff/6$ are not identically zero. To see this, note that it vanishes for a vacuum energy source, $\tau_{\mu\nu}=-\delta V_{vac}\bar g_{\mu\nu}$. This is exactly as it should be given the ability of the theory to self-tune in these limits. 

At the other extreme, for short wavelength sources like planets and stars, we assume that $\Bsq \gg |q^2|, |\meff|, |m^2_\phi|$, so that the amplitude reduces to 
\begin{equation}
\label{AmpUV}
    \mathrm{Amp} \sim -\frac{1}{\BA^2} \int \mathrm{d}^4x\sqrt{-\bar g} \bigg\{ \tau'^{\mu \nu} \frac{1}{\Bsq } \tau_{\mu \nu}  
     - \frac{1}{2} \tau ' \frac{1}{\Bsq } \tau + \frac{\epsilon}{2} \tau ' \frac{1}{\Bsq } \tau \bigg\} ,
\end{equation}
where 
$
\epsilon=\frac{m^2_\phi+2\meff}{6 \Gamma}
$.
If we identify $\BA^2$ with $M_{pl}^2/2$, the first two terms in \eqref{AmpUV} recover the amplitude for four dimensional GR, on flat space.  The last term represents the deviation away from GR, which is constrained to be small on solar system scales, such that $|\epsilon| \lesssim 10^{-5}$ \cite{will2014confrontation}.

Bringing all of this together, we see that there are three scenarios of interest. 
\begin{enumerate}
\item\emph{massive graviton, no tachyons, AdS vacuum}:  If $m_g^2, m_\phi^2 >0$, the constraints above require an AdS background and masses that also satisfy $\meff \lesssim \mathcal{O}(10^{-5}) q^2$ and $m_\phi^2 \lesssim \mathcal{O}(10^{-5}) q^2$.  Since we further assume that $q^2 \lesssim H_0^2$ this corresponds to an ultralight graviton and an ultralight scalar. The latter is also  very weakly coupled to matter, being five orders of magnitude more weakly coupled than the graviton.  
\item\emph{massless graviton, no tachyons, AdS vacuum}:  If $m_g^2=0, m_\phi^2 >0$, the constraints  require an AdS background and a scalar mass that also satisfies $m_\phi^2 \lesssim \mathcal{O}(10^{-5}) q^2$.  Since we assume that $q^2 \lesssim H_0^2$ this, once again, corresponds to an ultralight scalar that is also  very weakly coupled to matter, being five orders of magnitude more weakly coupled than the graviton.  
\item\emph{massless graviton, tachyonic scalar, dS vacuum}:   If $m_g^2=0, m_\phi^2 <0$, the constraints require a dS background and a scalar mass that satisfies $0>m_\phi^2 \gtrsim -\mathcal{O}(10^{-5}) |q^2|$. Since we assume that $|q^2| \lesssim H_0^2$ this corresponds to an ultralight tachyon with a lifetime that exceeds the current age of the universe. The scalar is also  very weakly coupled to matter, being five orders of magnitude more weakly coupled than the graviton.  

\end{enumerate}
In each of these scenarios, the theory exhibits self-tuning and is ghost-free. The third and final scenario contains a tachyonic instability, although it is very slow. Compatibility with solar system tests of gravity is achieved in each case thanks to the very weak coupling of the light scalar field. 
 
 \section{The canonical example} \label{sec:can}
 Let us now consider a canonical example with a massless graviton, $\bar{\mathcal{A}}^2 = M_{pl}^2/2 $, $\mathcal{K}=1/2$ and Ringo potential $V_r = \phi / \mu$, given by the action \eqref{canact}, only now we allow for the fact that the vacuum can be either dS or AdS, depending on the sign of $q^2$ (in our notation, the vacuum has curvature $-q^2$, being dS for $q^2<0$ and  AdS for $q^2>0$). As stated previously,  the vacuum energy $V_{vac} $ does not affect the vacuum curvature. Instead, it fixes the constant vev of the scalar field as $\bar \phi=-\frac{\mu}{24 q^4} (  V_{vac}+3 M_{pl}^2 q^2)$. 
 
 The exchange amplitude between conserved  sources is given by
\begin{multline}
\label{Ampc}
    \mathrm{Amp} = -\frac{2}{M_{pl}^2} \int \mathrm{d}^4x \sqrt{-\bar g} \bigg\{ \tau'^{\mu \nu} \frac{1}{\Bsq +2q^2 } \tau_{\mu \nu} 
     - \frac{\tau '}{2} \bigg[ \frac{1/2}{\Bsq +2q^2 } + \frac{1/2}{\Bsq -6q^2} \bigg] \tau
    + g_\phi \tau ' \frac{1}{\Bsq -  m^2_\phi} \tau \bigg\} ,
\end{multline}
where $$m^2_\phi =\frac{384 q^6}{M_{pl}^2\mu^2+96 q^4}, \qquad g_\phi=\frac{16 q^4}{M_{pl}^2\mu^2+96 q^4}.$$
At large wavelengths,  $\Bsq \ll q^2, m_\phi^2$, there is an elegant cancellation between terms and the amplitude acquires the tensor structure associated with self-tuning
\begin{equation}
\label{Ampa}
    \mathrm{Amp} \sim  -\frac{2}{M_{pl}^2} \int \mathrm{d}^4x \sqrt{-\bar g} \frac{1}{2 q^2 } \bigg\{ \tau'^{\mu \nu}  \tau_{\mu \nu} 
     - \frac14 \tau ' \tau \bigg\} .
\end{equation}
Note the importance of the background curvature $q^2 \neq 0$ for this to be well defined. At short wavelengths, $\Bsq \gg q^2, m_\phi^2$, the amplitude scales as
\begin{equation}
\label{Ampb}
    \mathrm{Amp} \sim -\frac{2}{M_{pl}^2} \int \mathrm{d}^4x \sqrt{-\bar g} \bigg\{ \tau'^{\mu \nu} \frac{1}{\Bsq  } \tau_{\mu \nu} 
     - \frac{1}{2} \tau ' \frac{1}{\Bsq } \tau
  + g_\phi \tau ' \frac{1}{\Bsq } \tau \bigg\} .
\end{equation}
If $|\mu| \gtrsim 1000 |q^2|/M_{pl}$, then $g_\phi \lesssim 10^{-5}$, ensuring compatibility with solar system tests of gravity. Further, the  size of the scalar mass $|m_\phi^2| \lesssim 10^{-4} |q^2|\lesssim 10^{-4} H_0^2$.  

\section{Discussion} \label{sec:dis}

In this paper, we have explored a family of generalised scalar-tensor theories, finding examples of self-tuning to low scale  AdS or dS vacua for any choice of vacuum energy.  Since translational invariance is broken by the background curvature, these self-tuning vacua can exist even for constant scalar profiles and still evade Weinberg's no go theorem.   This makes it easy to investigate fluctuations around the vacuum, which can be made compatible with solar system gravity tests provided the scalar is very weakly coupled. 

We should not be surprised that compatibility with solar system tests required a weakly coupled scalar.  Such a statement is trivial if the scalar exists alongside a massless graviton. It is also true for an ultra light massive graviton in anti de Sitter space, thanks to the absence of the so-called vDVZ discontinuity in that case \cite{adsvdvz}.  What is, perhaps, more surprising, is that self-tuning can be achieved even when the scalar is so weakly coupled.  However, the point is that the scalar is also ultra light. This can compensate the effect of weak coupling in the presence of a vacuum energy source, giving a non-trivial contribution to the amplitude.  

The weak coupling should also help protect the stability of these theories under radiative corrections.  We can see this  explicitly in the canonical model \eqref{canact}.  Here there is a shift symmetry in the scalar broken only by the linear potential.  The coefficient of the linear potential, $q^4/\mu \lesssim M_{pl} q^2/1000$, sets the scale of the background vacuum and is technically natural. Note that the scale $\mu$ is only very weakly constrained.

The weak  scalar coupling could have some important (perhaps worrying) consequences for early universe phase transitions, in which the vacuum energy jumps over a relatively short timescale.  Suppose a transition occurs when $H \approx H_*$, at which point  vacuum energy jumps as $\Delta V_{vac} \sim M_{pl}^2H_*^2$. This will trigger a change in the scalar $\Delta \phi$, as it moves from the self-tuning position prior to the transition, to the one after.  If $g_\phi$ measures the strength of the scalar coupling,  we might naively expect the scalar to take a proper time $\Delta t \sim g_\phi^{-1} H_*^{-1}$ to find the new self-tuning position after the transition.  Given that $g_\phi \lesssim 10^{-5}$, this is likely to take more than 100,000 Hubble times, during which the vacuum energy is expected to gravitate, leading to a long period of inflation. The universe will cool significantly during this time, as the matter content of the universe is rapidly diluted.  This sounds problematic, especially when we consider the fact that the QCD phase transition occurs at a low scale $\Lambda_{QCD} \sim 200$ MeV.  As a result, there are likely to be significant phenomenological challenges that must be overcome for these self-tuning models to be considered viable in the presence of vacuum energy phase transitions. 

Of course, it is important to establish whether or not our  naive expectation is indeed realised, and this is a computationally non-trivial task. The details of a dynamical response to a phase transition could also depend on the inclusion of  Paul or  John terms in the underlying action. These terms do not affect the perturbative analysis presented here.  Assuming the response to the QCD transition is indeed slow, could we identify the burst of accelerated expansion as a model for early universe inflation? In this set-up, the self-tuning scalar would be identified with the inflaton, and inflation would be halted once it had found the new self-tuning vacuum. Of course, the universe would need to warm up again, raising the temperature beyond the scale of nucleosynthesis, but that is not in general a problem for preheating, which would probably be the preferred route to reheat the universe because the scalar and matter are only very weakly coupled.

\acknowledgements We would like to thank Claudia de Rham for useful conversations and Liam O’Brien for collaboration at an early stage of the project.
The work of AP and EJC was funded by STFC Consolidated Grant Number ST/T000732/1. The work of SG was funded by an STFC studentship.

\appendix 
 
\section{Exchange Amplitude}
\label{Exchange Amplitude}
Here we derive the formula for the amplitude \eqref{Amp}. We begin by perturbing equations \eqref{ScalarEF} and \eqref{EFE} about the background solution with $g_{\mu \nu} = \Bar{g}_{\mu \nu} + h_{\mu \nu}$, $\phi = \Bar{\phi} +\delta \phi$, $T^\mu_\nu =-V_{vac} \delta^\mu_\nu + \tau^\mu_\nu$. The perturbed field equations are given as $\delta \Ec^\mu_\nu = \delta \Ec_\phi = 0$ where
    \begin{equation}
    \label{PerturbG}
        \delta \Ec^\mu_\nu = - 2 \BA^2 \delta G^\mu_\nu + (4\BA \BA' - 8q^2 \BRingo ')D^\mu_\nu \delta \phi 
        + \meff \big( 6\BA \BA ' \delta \phi \delta^\mu_\nu+ \BA^2 \big( h\delta^\mu_\nu - h^\mu_\nu\big) \big) 
         + \tau^\mu_\nu ,
    \end{equation}
    and
    \begin{equation}
    \label{PerturbScalar}
        \delta \Ec_\phi = 2\BA \BA ' \delta R + 3\meff \BA \BA ' h + 2\BK \Bsq \delta \phi + \BRingo ' \big( 4 \Bar{P}_{\mu}{}^\alpha {}_{\nu \beta}\Bnab_\alpha \Bnab^\beta h^{\mu \nu} - 12q^4 h \big) .
    \end{equation}
 Here  $D^\mu_\nu \equiv \Bnab^\mu \Bnab_\nu - \Bsq \delta^\mu_\nu + 3q^2 \delta^\mu_\nu$ and $\meff \equiv m^2 (3\BA -2)$, and indices are raised and lowered with respect to the background metric. 

We now decompose $h_{\mu \nu}$ as
\begin{equation}
\label{hdecompose}
   h_{\mu \nu} = h_{\mu \nu}^{(TT)} + 2\Bnab_{( \mu} A^{(T)}_{\nu )}+2\Bnab_\mu \Bnab_\nu \chi+ 2 \Bar{g}_{\mu \nu}\psi ,
\end{equation}
where $ A^{(T)}_\mu$ is transverse  and $ h_{\mu \nu}^{(TT)}$ is transverse-tracefree
$$
\Bnab^\mu A^{(T)}_\mu=0, \qquad \bar g^{\mu\nu}h_{\mu \nu}^{(TT)} =0=\Bnab^\mu  h_{\mu \nu}^{(TT)} .
$$
We can consistently set the vector part $A_\mu^{(T)}$ to $0$. By using the Lichnerowicz operator to compute $\delta G^\mu_\nu$, equation \eqref{PerturbG} can now be written as 
\begin{equation}
    \label{tauTT2}
        \tau^{(TT)}{}^\mu_\nu = -\BA^2 \big( \Bsq + 2q^2 - \meff \big) h^{(TT)}{}^\mu_\nu,
    \end{equation}
    where the transverse-tracefree part of the energy-momentum tensor is given by
    \begin{equation}
    \label{tauTT}
        \tau^{(TT)}{}^\mu_\nu = \tau^\mu_\nu-2D^\mu_\nu \mathcal{S}+C\delta^\mu_\nu ,
        \end{equation}
    with
        \begin{equation} \label{sig}
\mathcal{S}=\BA^2 \big(\meff \chi - 2  \psi  -2\alpha \delta \phi \big) +4q^2 \BRingo '  \delta \phi ,
\end{equation}
and 
\begin{equation}
\label{constraint}
    C = 6\BA^2\meff  \big( \psi +q^2 \chi +\alpha \delta \phi \big) .
\end{equation}
Here we have also defined $\alpha \equiv \frac{\BA '}{\BA}$. Assuming conservation of the source   $\Bnab_\mu\tau^\mu_\nu =0$, the transverse property of \eqref{tauTT} implies  the  constraint $C=0$. Note that this only gives something non-trivial if $\meff \neq 0$, when the massive graviton breaks diffeomorphisms explicitly. 

The transverse-tracefree part of the energy momentum tensor can now be written more succinctly as
$
        \tau^{(TT)}{}^\mu_\nu = \tau^\mu_\nu -2D^\mu_\nu \mathcal{S}
$,  where we now have
\begin{equation} \label{sig1}
\mathcal{S}=-\frac{1}{q^2} \left(\BA^2 \Gamma \psi+\mathcal{R}\delta \phi\right),
\end{equation}
with $\Gamma \equiv 2q^2 + \meff$ and $\mathcal{R}=\alpha \Gamma \BA^2  -4q^4 \BRingo' $. Note that we have also used the constraint $C=0$ to eliminate $\meff \chi$ in our expression for $\mathcal{S}$.

Using the tracelessness of $\tau^{(TT)}{}^\mu_\nu$, we further obtain
\begin{equation} \label{sig2}
\mathcal{S}=-\frac16 \frac{1}{\Bsq-4q^2} \tau .
\end{equation}
An explicit computation of the perturbed scalar equation of motion 
 \eqref{PerturbScalar} now yields
 \begin{equation}  \label{pphieq}
 -6\frac{\mathcal{R}}{q^2} (\Bsq-4q^2)\psi+2\mathcal{K}_\text{eff} \Bsq
 \delta \phi=0 ,
 \end{equation}
 where $\mathcal{K}_\text{eff} =\BK-\frac{3\alpha^2 \meff\BA^2}{q^2}$,  and, once more, we have used the constraint $C=0$ to eliminate $\meff \chi$ in our final expression. We can use equation \eqref{sig1} to eliminate $\delta \phi$ in favour of $\mathcal{S}$. Plugging the result into equation \eqref{pphieq} we obtain
 \begin{equation} \label{psisol}
 \psi=\frac{m_\phi^2-4q^2}{4\BA^2 \Gamma} \frac{\Bsq}{\Bsq-m_\phi^2}\mathcal{S} ,
 \end{equation}
 where $m_\phi^2$ is given by equation \eqref{mphi}. 

We are now ready to compute the exchange amplitude for two conserved sources, $\tau_{\mu \nu}$ and $\tau_{\mu \nu}'$, on a maximally symmetric background, with curvature $-q^2$. To do this we write
\begin{equation}
    \begin{split}
        \mathrm{Amp} 
        & = \int \mathrm{d}^4x \sqrt{-\bar g} \tau '^{\mu \nu} h_{\mu \nu}   = \int \mathrm{d}^4x  \sqrt{-\bar g}\big( \tau '^{\mu \nu} h^{(TT)}_{\mu \nu} + 2 \tau ' \psi \big) ,
    \end{split}
\end{equation}
where we have used \cref{hdecompose}, along with energy momentum conservation to eliminate  the $\Bnab_\mu \Bnab_\nu \chi$ terms.  Using the fact that 
\begin{equation}
h^{(TT)}_{\mu \nu} =-\frac{1}{\BA^2} \frac{1}{\Bsq +2q^2 -\meff} \left(\tau_{\mu\nu} -2 D_{\mu\nu}\mathcal{S}\right) ,
\end{equation}
and our formula for $\psi$, given by equation \eqref{psisol}, the amplitude can be written as 
    \begin{multline}
        \mathrm{Amp} = -\frac{1}{\BA^2} \int \mathrm{d}^4x \sqrt{-\bar g}   \bigg\{ \tau'^{\mu \nu} \frac{1}{\Bsq +2q^2 -\meff} \tau_{\mu \nu} 
        -2 \tau'^{\mu \nu} \frac{1}{\Bsq +2q^2 -\meff} D_{\mu \nu} \mathcal{S} \\
        - \tau'  \frac{m_\phi^2-4q^2}{2\Gamma} \frac{\Bsq}{\Bsq-m_\phi^2}\mathcal{S} \bigg\} .
    \end{multline}
We now make use of the following formula  \cite{niedermann2017gravitational}
\begin{equation}
\frac{1}{\Bsq-M^2} \bar \nabla_{\mu} \bar \nabla_{\nu} \mathcal{S}= \bar \nabla_{\mu} \bar \nabla_{\nu} \frac{1}{\Bsq-M^2-8q^2} \mathcal{S} +\frac14 \bar g_{\mu\nu} \left[\frac{M^2}{ \Bsq-M^2}-\frac{(M^2+8q^2)}{\Bsq-M^2-8q^2}\right]\mathcal{S} ,
\end{equation}
and energy-momentum conservation, so that the amplitude becomes
\begin{widetext}
   \begin{multline}
        \mathrm{Amp} = -\frac{1}{\BA^2} \int \mathrm{d}^4x \sqrt{-\bar g}   \bigg\{ \tau'^{\mu \nu} \frac{1}{\Bsq +2q^2 -\meff} \tau_{\mu \nu} 
        +2 \tau' \frac{\Bsq-3q^2}{\Bsq +2q^2 -\meff}\mathcal{S} \\
         - \frac12 \tau' \left[\frac{-2q^2 +\meff}{ \Bsq+2q^2 -\meff}-\frac{(6q^2 +\meff)}{\Bsq-6q^2-\meff}\right] \mathcal{S}
        -  \tau'  \frac{m_\phi^2-4q^2}{2\Gamma} \frac{\Bsq}{\Bsq-m_\phi^2}\mathcal{S} \bigg\} .
    \end{multline}
    \end{widetext}
Upon inserting the form of $\mathcal{S}$ given by equation \eqref{sig2},  and simplifying, we arrive at the formula for the amplitude stated in equation \eqref{Amp}.
\bibliography{MassiveGAdS}
\bibliographystyle{unsrt}
\end{document}